\documentclass{ws-p8-50x6-00}

\def\etal{{\it et al.}\ }
\def\Mpc{$h_{100}^{-1}$~{\rm Mpc}}
\def\hmpc{$h$~{\rm Mpc$^{-1}$}}

\begin{document}

\title{Large Scale Structure of the Universe: Current Problems}

\author{J. Einasto}

\address{Tartu Observatory, EE-61602 T\~oravere, Estonia}

\maketitle

\abstracts{
I compare the mean power spectrum of galaxies with theoretical models
and discuss possibilities to explain the observed power spectrum.  My
principal conclusion is that some of the presently accepted
cosmological paradigms need revision if the available observational
data represent a fair sample of the Universe.  }

\section{Introduction}

According to current paradigms the Universe is homogeneous and
isotropic on large scales, density perturbations grow from small
random fluctuations generated in the early stage of the evolution
(inflation), and the dynamics of the Universe is dominated by cold
dark matter (CDM) with some possible mixture of hot dark matter
(HDM). On small scales galaxies are associated in groups and
clusters. Until recently it was assumed that the homogeneity of the
Universe starts on scales above 50~\Mpc.  However, there is growing
evidence that the supercluster-void network has some regularity, and
that homogeneity occurs on larger scales only.  Broadhurst \etal
(1990) measured redshifts of galaxies in a narrow beam towards the
northern and southern Galactic poles and found that the distribution
is periodic: high-density regions (which indicate superclusters of
galaxies, see Bahcall 1991) alternate with low-density ones (voids)
with a surprisingly constant interval of $\approx 128$~\Mpc\ (here $h$
is the Hubble constant in units of 100~km~s$^{-1}$~Mpc$^{-1}$).  The
distribution of galaxies in this beam can be characterized by a power
spectrum which has a sharp peak. The power spectra derived on the
basis of large cluster samples have maxima around the same scale
(Einasto \etal 1997a, 1999a). Below we shall analyze observed power
spectra of galaxies and clusters of galaxies and compare them with
theoretical spectra found for various models.

\section{The power spectrum of galaxies and its interpretation}

\begin{figure}
\vspace{5.0cm}
\caption{The left panel  shows empirical power spectra of matter.
$P_{n.lin}$ is the non-linear power spectrum of matter with its
$1\sigma$ error corridor; $P_{lin}$ and $P_{APM-lin}$ are the linear
power spectra for empirical spectra derived from large cluster samples
and from the APM 2-D galaxy sample (Peacock 1997, Gazta\~naga \& Baugh
1998), respectively. The right panel gives the empirical linear power
spectrum of matter compared with theoretical spectra for MDM models
with $\Omega_{0}=1.0, ~0.9, \dots ~0.4, ~0.25$; for clarity the models
with $\Omega_0 = 1.0$ and $\Omega_0 = 0.5$ are drawn as dashed
lines.  }
\includegraphics{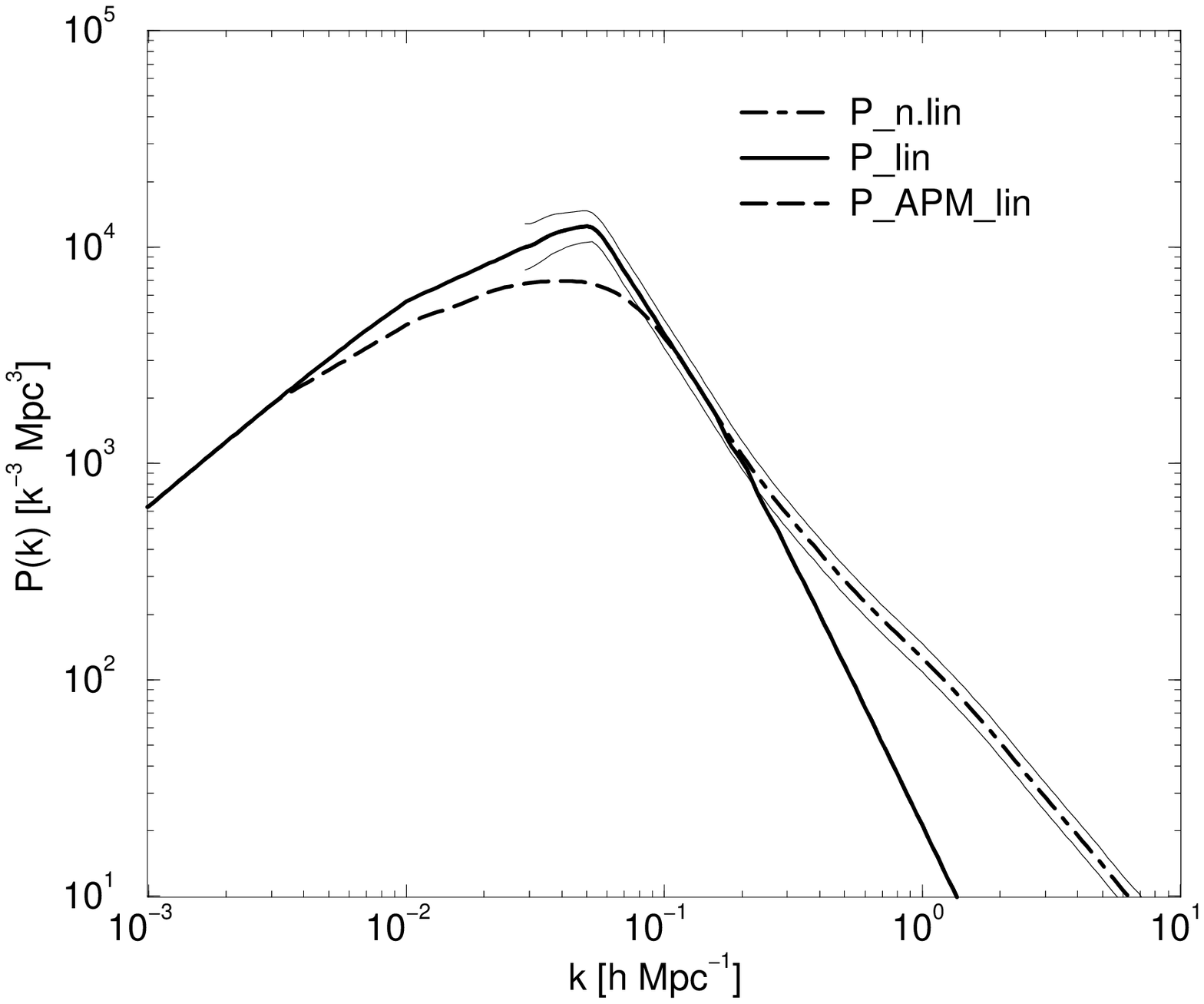}
\includegraphics{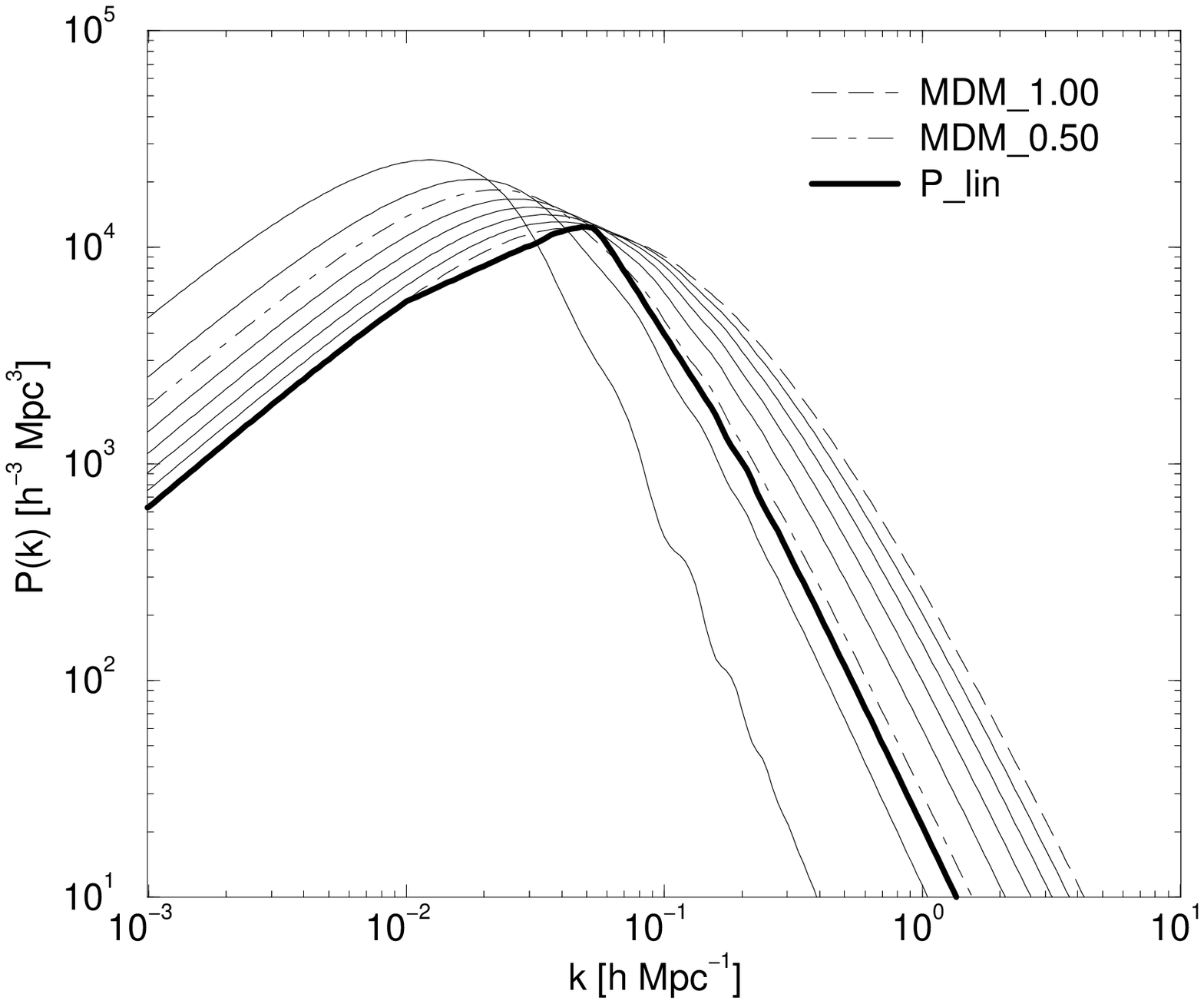}
\end{figure}

We have analyzed recent determinations of power spectra of large
galaxy and cluster samples.  The mean power spectrum found from
cluster samples (Einasto \etal 1997c, Retzlaff \etal 1998, Tadros
\etal 1998) and the APM 3-D galaxy sample (Tadros and Efstathiou 1996)
has a relatively sharp maximum at wavenumber $k= 0.05$~\hmpc, which
corresponds to a scale of 120~\Mpc, and an almost exact power law with
index $n=-1.9$ on scales shorter than the maximum.  In contrast, the
power spectrum found from deprojection of the 2-D distribution of APM
galaxies (Peacock 1997, Gazta\~naga \& Baugh 1997) is shallower around
the maximum, see Figure~1.  We may expect that true 3-D and deeper
surveys reflect better the actual distribution of galaxies and
clusters, thus we assume that the power spectrum based on cluster data
is characteristic for all galaxies in a fair sample of the Universe
(Einasto \etal 1999a,b).  The spectrum is derived in real space, then
reduced to the amplitude of the spectrum of matter, and finally
corrected for non-linear effects; it is determined from observations
on scales $\leq 200$~\Mpc, while on very large scales it is
extrapolated using theoretical model spectra.

\begin{figure}
\vspace{5.5cm}
\caption{ The empirical linear power spectrum of matter compared with
theoretical spectra for MDM models of variable baryon density. Left
panel shows model spectra with zero vacuum energy density, right panel
with $\Omega_{\Lambda} = 0.6$. 
 }  
\includegraphics{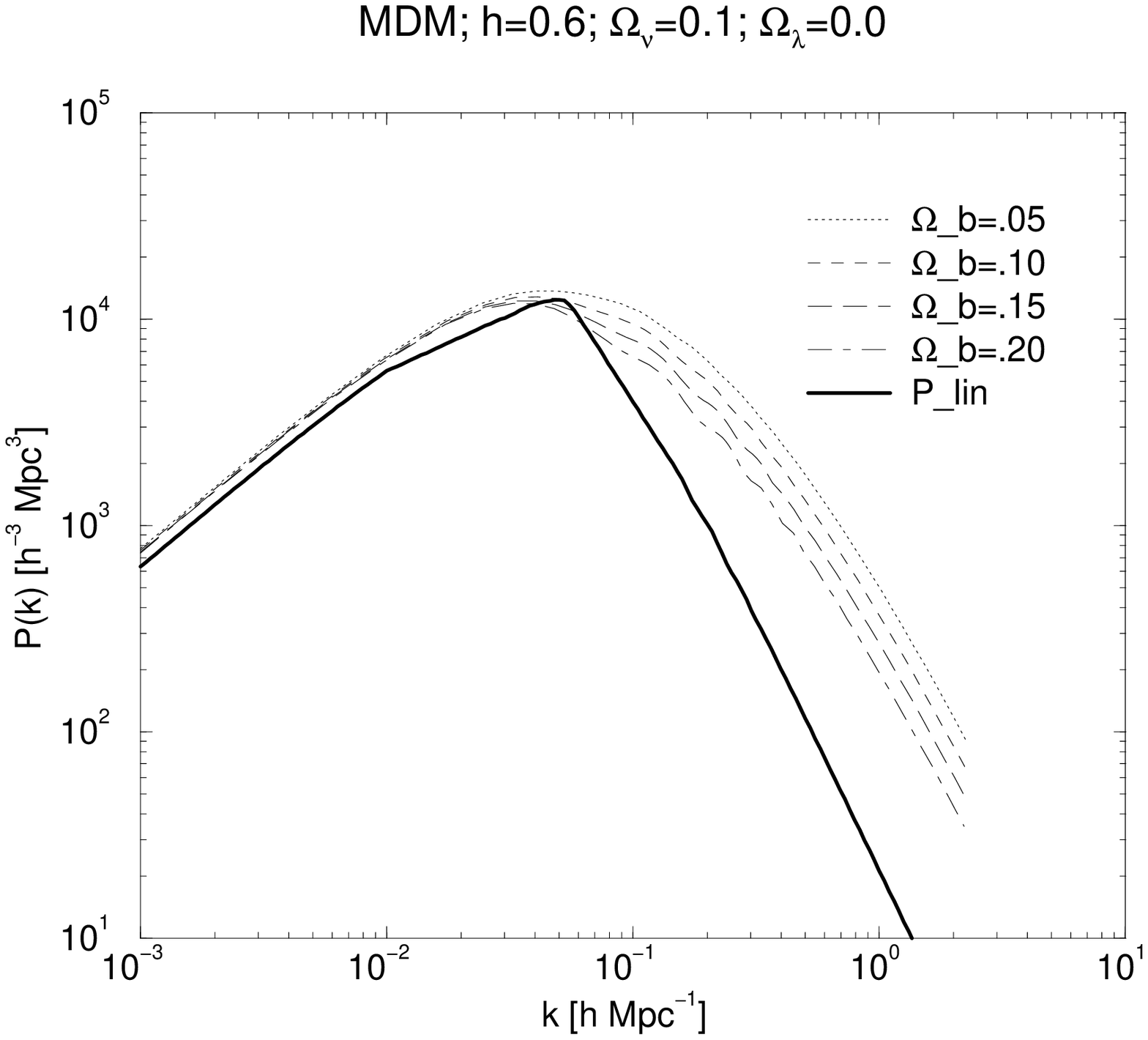}
\includegraphics{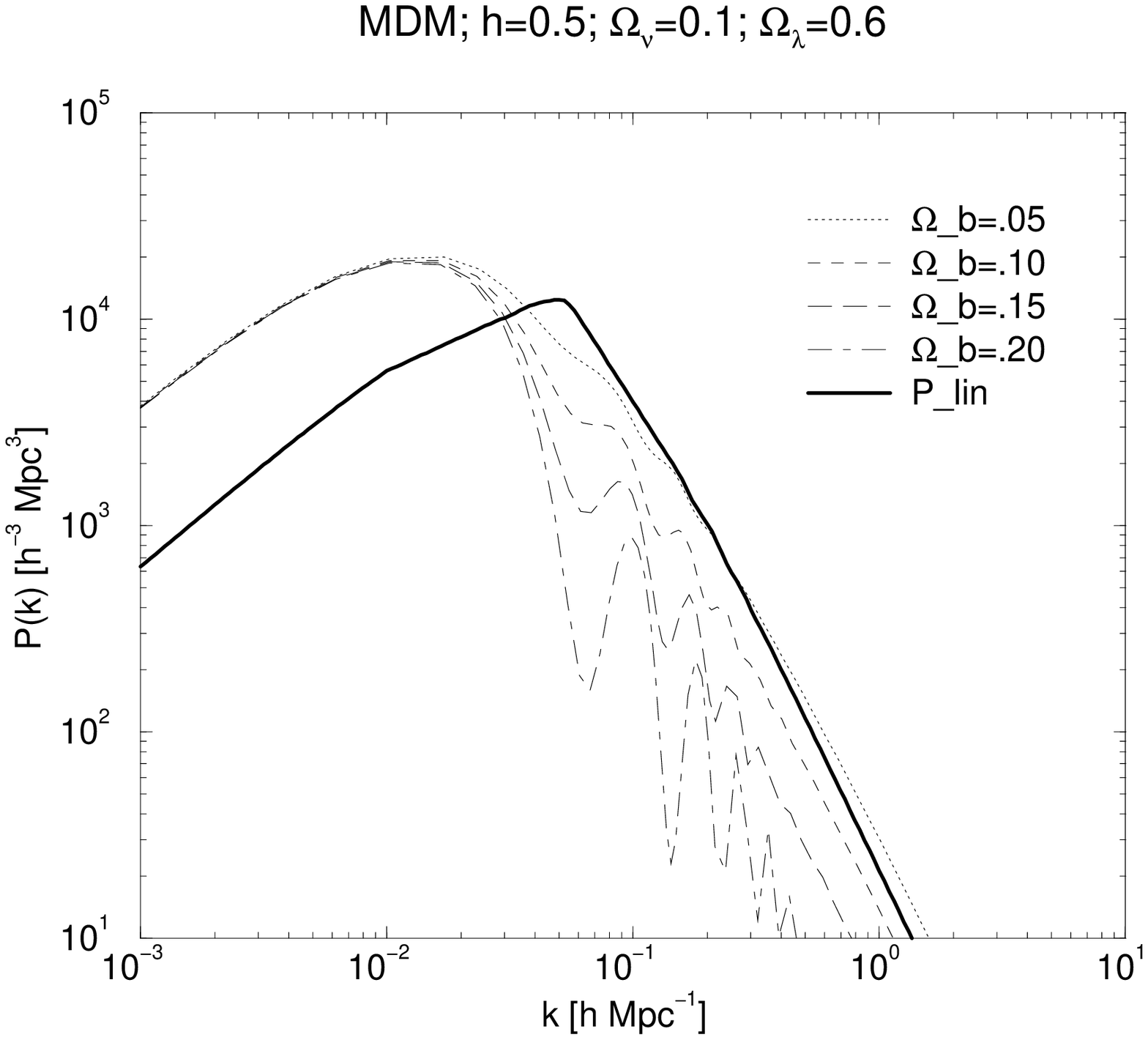}
\end{figure}

In the right panel of Figure~1 we compare the empirical power spectrum
with theoretical models. The best agreement is achieved with a mixed
dark matter (MDM) model with cosmological constant. We have accepted
parameters of models in agreement with recent data: Hubble constant
$h=0.6$; baryon density $\Omega_b=0.04$ (this gives $\Omega_b h^2 =
0.0144$); hot dark matter density $\Omega_{\nu}=0.1$.  We use
spatially flat models with cosmological constant, $\Omega_0 +
\Omega_{\Lambda} = 1$. The matter density $\Omega_0=\Omega_b +
\Omega_{c} + \Omega_{\nu}$ was varied between 1.0 and 0.25, the cold
dark matter fraction $\Omega_{c}$ and vacuum energy term
$\Omega_{\Lambda}$ were chosen in agreement with restrictions given
above.  The amplitude of power spectra on large scales was normalized
using four-year COBE data.  All models are based on the basic
assumption that the primordial power spectrum is a power law; we have
calculated model spectra for power indices of $1.0,~1.1, \dots~1.4$;
model spectra plotted in Figure~1 were derived for $n=1.0$.

Figure~1 shows that on scales shorter than the scale of the maximum
the best agreement with observations is obtained with a model with
density parameter $\Omega_0 \approx 0.4$.  By fine tuning the density
parameter $\Omega_0$ and power index $n$ it is possible to get an
almost exact representation of the empirical power spectrum on scales
$<120$~\Mpc.  The agreement is lost on large scales.  The power
spectra of models with low density value continue to rise toward large
scales as seen in Figure~1.  An agreement with the amplitude of
theoretical power spectra is possible only for models with high
density parameter, $\Omega_0 \approx 1$.  However, models with high
density parameter have amplitudes much higher than the empirical
spectrum.  It is impossible to satisfy the shape of the empirical
power spectrum with models with any fixed density parameter
simultaneously on large and small scales.  This is the main conclusion
obtained from the comparison of cosmological models with the data.
The main reason for this disagreement is that the empirical power
spectrum is narrower: its full width at half height of the maximum is
about 0.8 dex, whereas conventional CDM models have this parameter in
the range $1.10 - 1.26$ dex, and MDM models in the range $1.00 - 1.15$
dex (lower value for lower density parameter).

\begin{figure}
\vspace{5.5cm}
\caption{  Distribution of clusters in high-density regions in
supergalactic coordinates. Left panel shows clusters in a sheet in   
supergalactic $-100 \leq X \leq 200$~\Mpc; Abell-ACO and APM clusters 
in superclusters with at least 8 or 4 members are plotted with symbols
as indicated.  The supergalactic $Y=0$ plane coincides almost exactly
with the Galactic equatorial plane and marks the Galactic zone of   
avoidance. In the right panel only clusters in the southern Galactic  
hemisphere are plotted; here the depth is $-350 \leq Y \leq -50$~\Mpc.
 }  
\includegraphics{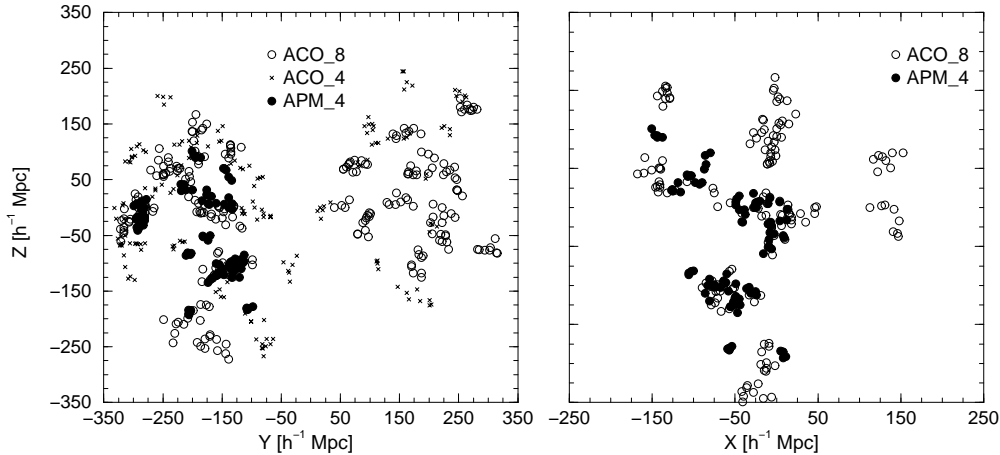}
\end{figure}

One possibility to explain this discrepancy and to decrease the width
of the power spectrum is to increase the baryon fraction in the cosmic
density budget (Eisenstein \etal (1998) and Meiksin, White \& Peacock
(1999)).  In this case the amplitude of Sakharov oscillations of the
hot plasma before recombination increases which decreases the width of
the power spectrum and increases the amplitude of the power spectrum
near the peak.  We have checked this possibility and calculated the
power spectra for a range of baryon densities, varying also the Hubble
parameter and vacuum energy density (cosmological constant term).
Figure~2 shows results for a set of MDM models with Hubble constant
$h=0.5$ and $h=0.6$, vacuum energy density $\Omega_{\Lambda} = 0.0$
and $0.6$, and HDM fraction $\Omega_{\nu} = 0.1$, the baryon density
was varied between 0.05 and 0.20.  The increase of the baryon fraction
in models with high cosmological density (and zero cosmological
constant) does not change the power spectrum considerably -- the width
of the spectrum remains too large.  In models with large cosmological
constant an increase of the baryon fraction decreases the width of the
power spectrum, however, Sakharov oscillations of the spectrum become
too large.  Moreover, the shape of all theoretical power spectra is
very different from the shape of empirical spectra.  The first peak of
Sakharov oscillations occurs on a scale of $k \approx 0.1$~\hmpc; the
location of the overall maximum of the power spectrum depends on
the density parameter. For low-density models it is located near $k
\approx 0.01$~\hmpc, the observed maximum lies in-between.  Varying
the Hubble constant does not change the overall picture, and there
remain essential differences between models and data.

Thus our calculations show that no combination of cosmological
parameters enables us to obtain a good representation of the empirical
power spectrum with theoretical models which are based on the
assumption that the primordial power spectrum is a single power law.
There remain two possibilities, either empirical data are in error or
the single power law assumption is wrong.

\section{Geometry of the distribution of clusters}

Consider first the possibility that the observed power spectrum is not
accurate enough, and that there is actually no discrepancy between
models and data.  Differences occur on scales near the maximum of the
spectrum.  Here density perturbations have the largest amplitudes,
thus it is clear that maxima correspond to superclusters --
large-scale regions of highest density in the Universe, and minima to
large voids between superclusters -- regions of lowest overall
density.  Differences in power spectra on these scales reflect
differences in the spatial distribution of superclusters and voids.
To understand the meaning of differences between observed and
theoretical power spectra we shall compare the distribution of real
and model superclusters and voids.  The most suitable objects to
investigate the distribution of superclusters are rich clusters of
galaxies.

Figure~3 presents the distribution of Abell-ACO and APM clusters of
galaxies located in rich superclusters with at least 4 or 8 member
clusters (Toomet \etal 1999).  To emphasize the distribution of
regions of highest density, which define the power spectrum near the
maximum, we plot only clusters in rich superclusters.  Figure~3 shows
that the distribution of rich clusters is quasi-regular: superclusters
and voids form a honeycomb-like pattern. The diameter of a cell in
this network is approximately 120~\Mpc, which is very close to the
scale of the maximum of the power spectrum.

In contrast to the observed case the distribution of rich
superclusters in CDM dominated models is almost random (Frisch \etal
1995).  Mock catalogues with randomly distributed superclusters have
power spectra with broad maxima similar to spectra of CDM-type models
(Einasto \etal 1997b).  The presence of broad maxima is an intrinsic
property of all CDM-type models (if the baryon fraction is not too
high). 

\begin{figure}
\vspace*{5.2cm}
\caption{ Power spectrum (left) and correlation function (right) of
the MDM  model with density parameter $\Omega_0=0.4$, compared
with the linear empirical power spectrum of matter and correlation
function of clusters of galaxies in rich superclusters. The power
spectrum of the MDM model is calculated with spectral index
$n=1.1$. Cluster correlation functions are calculated via Fourier
transform from power spectra of matter, and are enhanced in amplitude
by a biasing factor 7.7 which corresponds to the mean difference
between respective power spectra. The observed cluster correlation
function is the one for Abell-ACO clusters in very rich superclusters
as derived by E97b.  }
\includegraphics{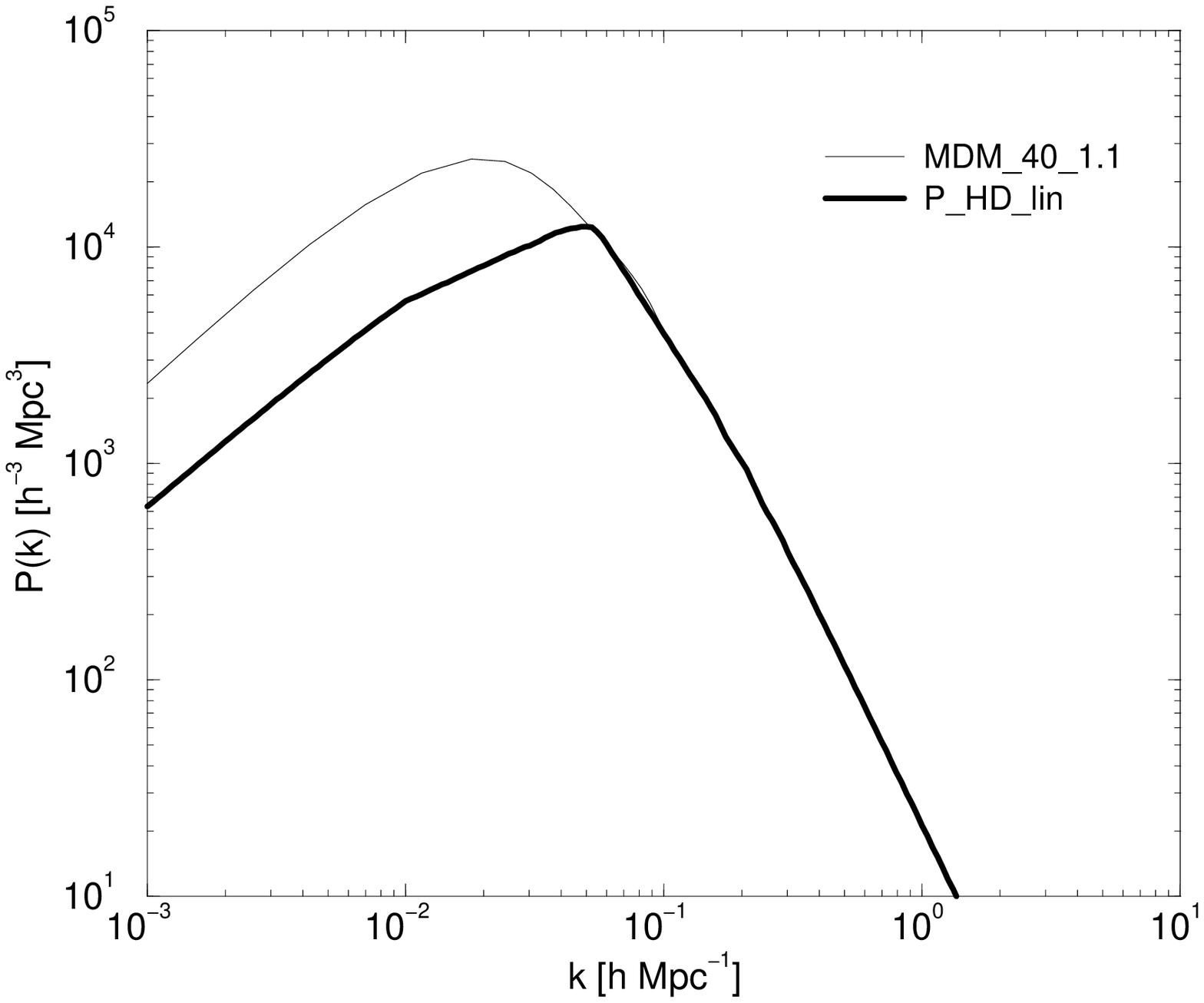}
\includegraphics{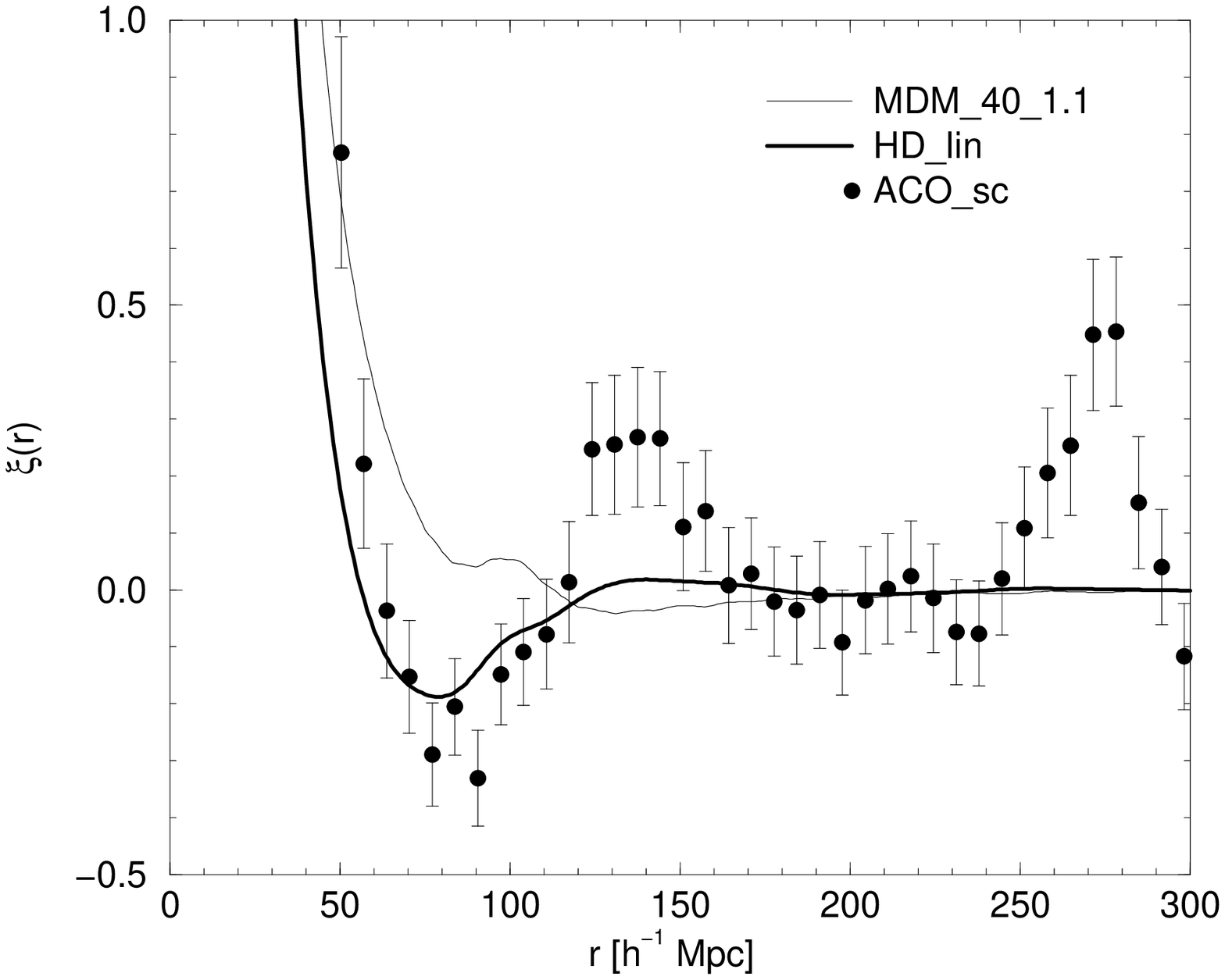}   
\end{figure}

The distribution of clusters can also be quantified using the
correlation function of clusters of galaxies.  While on small scales
the correlation function characterizes the distribution of clusters
within superclusters, on large scales it describes the distribution of
superclusters themselves (Einasto \etal 1997a, 1999b).  In Figure~4 we
compare power spectra and correlations functions of the MDM model for
a density parameter $\Omega_0=0.4$ with respective empirical data.  We
use the observed correlation function of clusters of galaxies located
in rich superclusters, and for comparison the Fourier transform of the
empirical power spectrum of matter, enhanced in amplitude to obtain a
correlation function comparable with the function for rich
superclusters.  The observed correlation function of clusters in rich
superclusters is oscillating with a period equal to the wavelength of
the maximum of the power spectrum.  The Fourier transform of the
empirical power spectrum has a similar property, only that the
amplitude of oscillations is lower.  The reason for this difference is
due to the elongated form of the cluster sample which enhances the
amplitude of oscillations at large separations.  These oscillations
are due to quasi-regular distribution of rich superclusters seen in
Figure~3.  The correlation function calculated for the MDM model has a
completely different character on large scales, and corresponds to an
almost random distribution of rich superclusters.

\begin{figure}
\vspace{5.2cm}
\caption
{Periodicity goodness curves for mock samples (left panel) and
Abell-ACO clusters (right panel). Mock samples have 300 randomly
located clusters and 100 clusters in quasi-regularly located
superclusters. The step of the regular grid is $r_0=130$~\Mpc; samples
are cubical. In both panels solid, dashed and dot-dashed lines are for
trial cubes oriented at $0^{\circ}$, $22.5^{\circ}$ and $45^{\circ}$
with respect to the main symmetry axis of the mock sample, and with
respect to supergalactic coordinates of the real sample. }
\includegraphics{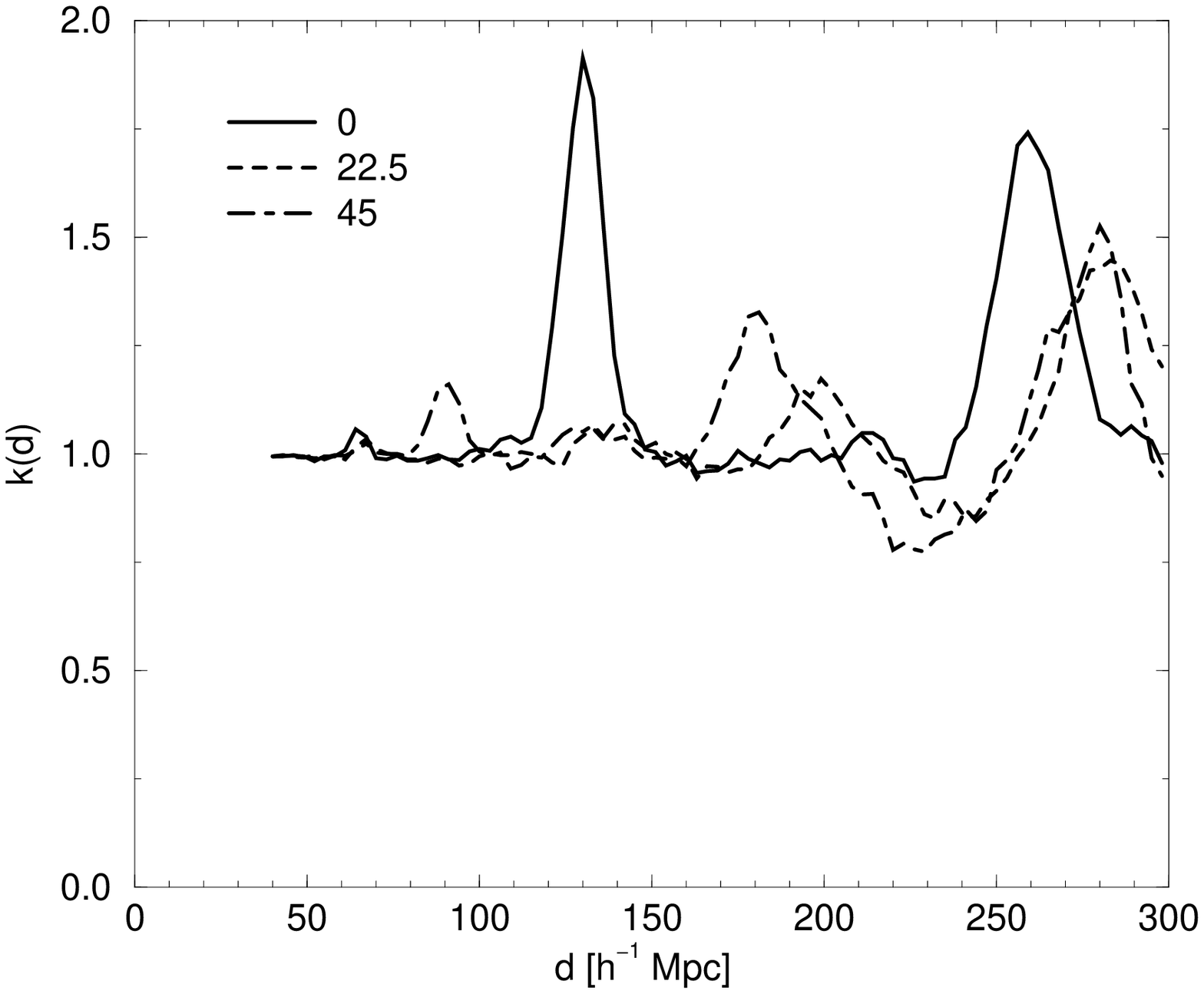}
\includegraphics{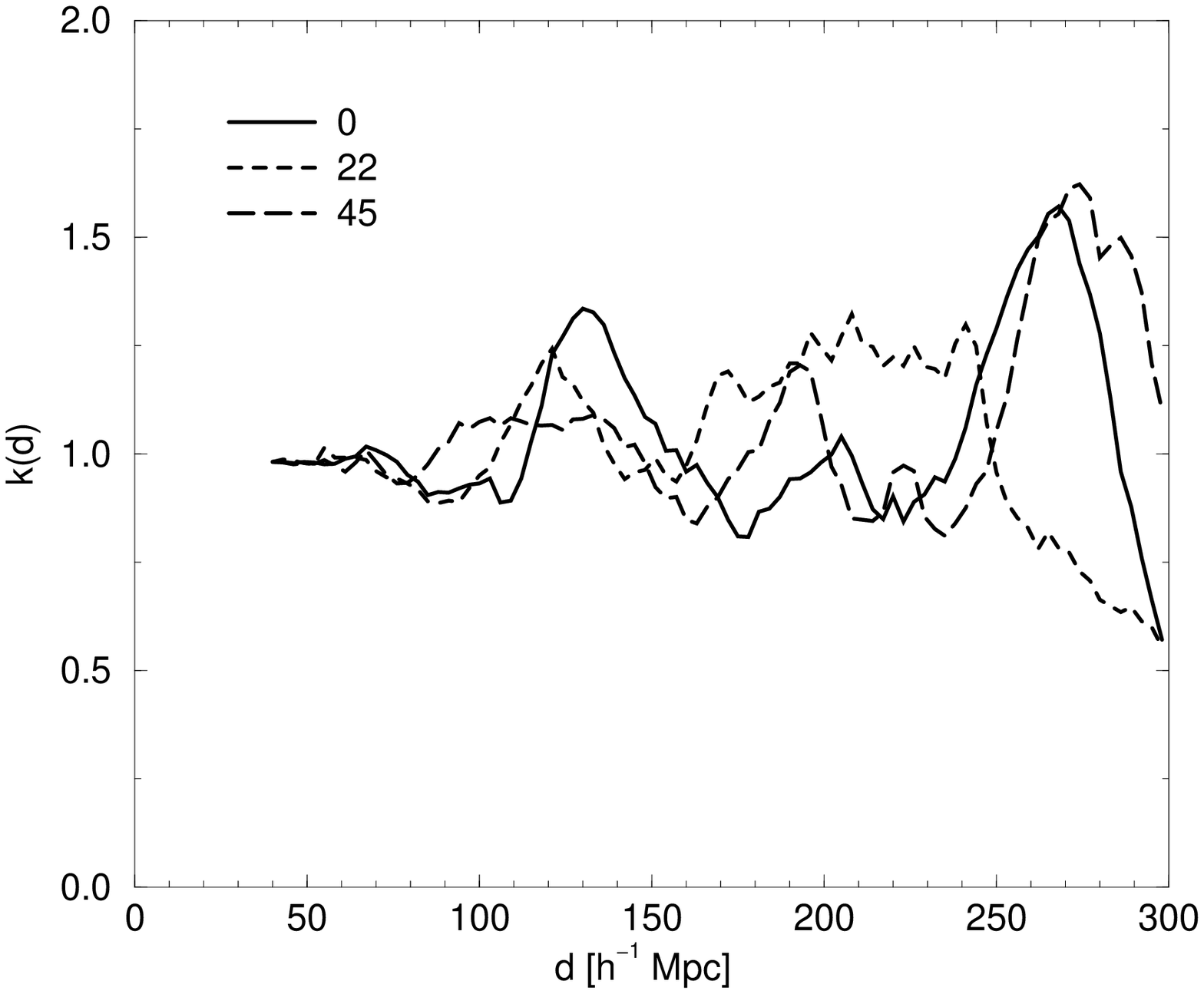}
\end{figure}

Finally, to describe the regularity of the cluster distribution we use
a novel method which is sensitive to the geometry of the distribution
(Toomet \etal 1999).  The method is a 3-D generalization of the
periodicity analysis of time series of variable stars.  The space
under study is divided into cubical trial cells of side length $d$.
All objects in the individual cubical cells are stacked to a single
combined cell, preserving their phases in the original cells. We then
vary the side-length of the trial cube to search for the periodicity
of the cluster distribution. We find the goodness of regularity for
the side length $d$ of the trial cell; it is defined so that it has a
maximum $>1$ if the length of the trial cell is equal to the period of
the regularity, otherwise it is equal to unity.  The goodness of
regularity is shown in Figure~5, the left panel gives results for a
mock catalogue (see Figure caption), the right panel for the actual
Abell-ACO cluster sample.

The method is sensitive to the direction of the axes of the trial
cubes.  If clusters form a quasi-rectangular cellular network, and the
search cube is oriented along the main axis of the network, then the
period is found to be equal to the side-length of the cell.  If the
search cube is oriented at some non-zero angle in respect to the major
axis of the network, then the presence of the periodicity and the
period depend on the angle.  If the angle is $\approx 45^{\circ}$,
then the period is equal to the length of the diagonal of the cell.
If the angle differs considerably from $0^{\circ}$ and $45^{\circ}$,
the periodicity is weak or absent.  As seen from Figure~5, the main
axis of the supercluster-void network is approximately oriented toward
supergalactic coordinates.  As the supergalactic $Y$ axis is very
close to the direction of the Galactic poles, it is natural to expect
a well defined periodicity in these directions as really observed by
Broadhurst \etal (1990). Our periodicity analysis confirms earlier
results on the presence of a high concentration of clusters and
superclusters towards both the Supergalactic Plane (Tully \etal 1992),
and towards the Dominant Supercluster Plane, which are oriented at
right angles with respect to each other (Einasto \etal 1997d).

\begin{figure}
\vspace{5.2cm}
\caption
{The primordial power spectra for MDM models; left for peaked
empirical power spectra, right for shallower spectra derived from APM
2-D galaxy data.  Primordial power spectra are divided by the
scale-free spectrum, $P(k) \sim k$.  Spectra are found for theoretical
transfer functions with $\Omega_{0}=1.0, ~0.9, \dots ~0.25$; for
clarity the spectra for models with $\Omega_0 = 1.0$ and $\Omega_0 =
0.5$ are drawn with dashed lines.
 }
\includegraphics{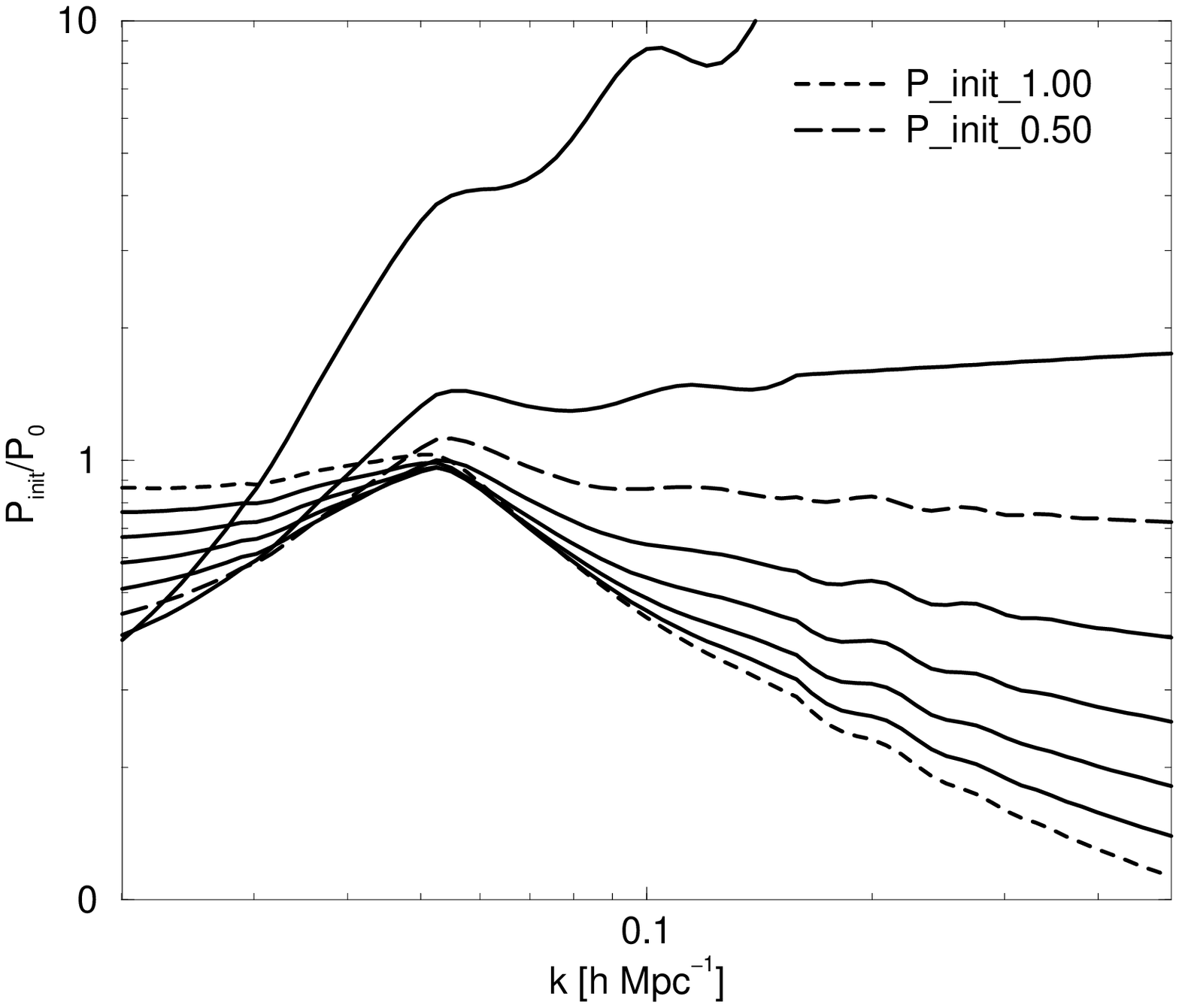}
\includegraphics{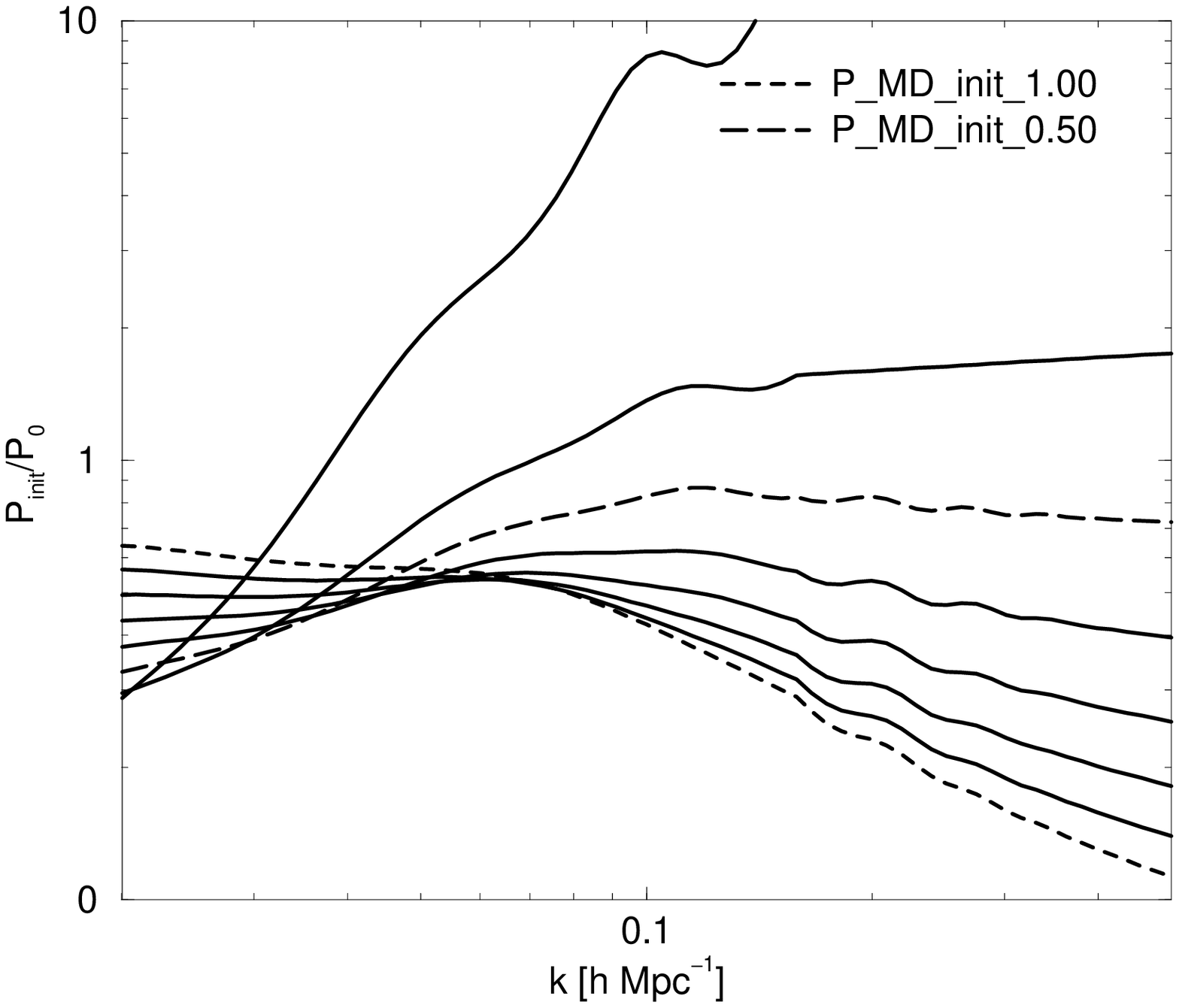}
\end{figure}

\section{Primordial power spectra}

Previous analysis has shown that there exist essential differences
between data and CDM-type models with scale-free primordial power
spectra.  To explain the difference between model and data we have to
accept a non-conventional theoretical power spectrum.  As we have
presently no reason to assume that our understanding of physical
processes during the radiation domination era is wrong, we suppose
that the peaked power spectrum originated during the earliest
inflational phase of the evolution of the Universe.  If we accept the
transfer function (which describes the evolution of the power spectrum
during the radiation domination era) according to models described
above, we derive the primordial power spectrum shown in Figure~6
(Einasto \etal 1999b).

The main features of primordial power spectra are the presence of a
spike and the change of the power index at the same scale as that of the
maximum of the empirical power spectrum.  On scales shorter or larger
than that of the spike, the primordial spectrum can be well
approximated by a power law.  The power indices of the approximation
are different on small and large scales.  Both alternative empirical
power spectra lead to similar primordial power spectra, only the shape
around the break is different.  Broken-scale-invariant primordial
power spectra have been studied by Starobinsky (1992), Adams, Ross \&
Sarkar (1997), and Lesgourgues \etal (1998), among others.  It is too
early to say which of these models describes the observational data
better.

\section{Conclusions}

Our main conclusions are:

{$\bullet$} The empirical power spectrum of matter has a peak on
scales near 120~\Mpc; on shorter scales it can be approximated by
a power law with index $n=-1.9$.

{$\bullet$} Superclusters and voids form a quasi-regular lattice
of mean cell size 120~\Mpc; the main axis of the lattice is directed
toward the supergalactic $Y$ coordinate.  
 
{$\bullet$} On scales around  100~\Mpc\ the Universe is 
neither homogeneous nor isotropic.

{$\bullet$} The primordial power spectrum of matter is broken,
its effective power index changes around the scale $\approx 120$~\Mpc.

I thank H. Andernach, F. Atrio-Barandela, M. Einasto, E. Kasak,
A. Knebe, V.  M\"uller, A. Starobinsky, E. Tago, O. Toomet and
D. Tucker for fruitful collaboration and permission to use our joint
results in this review article.  This study was supported by the
Estonian Science Foundation.

\end{document}